\documentclass[a4paper]{article} 
\usepackage[a4paper]{geometry} 
\usepackage{amsmath}
\usepackage{amssymb}
\usepackage{paralist} 

\usepackage{layouts} 
\usepackage{float} 
\usepackage{authblk} 

\usepackage{caption}
\usepackage{color}
\usepackage{graphicx}

\usepackage[figuresonly,nomarkers,nofiglist,notablelist]{endfloat}
 

\renewcommand*\familydefault{\sfdefault}

\title{On the precision of neural computation with interaural time
differences in the medial superior olive}

\def\Pathologic{a}
\def\Radio{b}
\def\Computer{d}
\def\Contact{c}
\def\Robotics{e}
\def\Poly{f}

\def\lb{, }
\def\rb{, }

\author[\Pathologic,\Radio,\Contact]{Petr Marsalek}
\author[\Computer]{Pavel Sanda}
\author[\Robotics,\Poly]{Zbynek Bures}

\affil[\Pathologic]{Institute of Pathological Physiology, First Medical
 Faculty, Charles University in Prague, U~Nemocnice~5/478,
 128~53, Praha~2, Czech Republic}

\affil[\Radio]{Department of Radioelectronics, Faculty of Electrical
 Engineering, Czech Technical University in Prague, Technic{k}\'a~2/1902, 
 166~27, Praha~6, Czech Republic}

\affil[\Computer]{Institute of Computer Science, Czech Academy
 of Sciences, Pod Vod\'{a}renskou v\v{e}\v{z}\'{\i}~2/271,
 182~07, Praha~8, Czech Republic}

\affil[\Robotics]{Czech Institute of Informatics, Robotics and Cybernetics,
 Czech Technical University in Prague,
 Jugosl\'{a}v{s}k\'y{c}h partyz\'an{\accent23u}~4/1903,
 166~36, Praha~6, Czech Republic}

\affil[\Poly]{College of Polytechnics,
 {T{o}l{s}t\'e{h}o}~16/1556, 586~01, Jihlava, Czech Republic}

\affil[\Contact]{{\it Corresponding author:} Petr.Marsalek@LF1.CUNI.CZ}

\long\def\comment#1{}

\begin{document}

\maketitle
  
\subsubsection*{Abstract}

{\small
Incoming sound is in cochlea and auditory nerve encoded into spike trains. At the third neuron of the auditory pathway, spike trains of the left and right sides are processed in brainstem nuclei to yield sound localization information.
Two different localization encoding mechanisms are employed in two ``centers'' for low and high sound frequencies in the brainstem. The centers are superior olivary nuclei, medial and lateral.
This paper contains analytical estimates of parameters needed in description of auditory coding in sound localization neural circuit.
%
Our model spike trains are based on electro-physiological recordings.
We arrive to best estimates for neuronal signaling with the use of just noticeable difference of the ideal observer. We describe spike timing jitter and its role in the spike train processing. We study frequency dependence of spike trains on the sound frequency. All parameters are accompanied with detailed estimates of their values and variability. Intervals bounding all the parameter from lower and higher values are discussed.
}


\subsubsection*{Keywords}
binaural hearing;
coincidence detection; ergodic hypothesis;
ideal observer; interaural time difference;
just noticeable difference; lateral and medial superior olive;
neuronal arithmetic; psychophysics; sound localization;
spike timing jitter.

\subsubsection*{Abbreviations and symbols}

$f_{\mathrm{S}}$,~sound frequency;
$F_{\mathrm{C}}$,~critical sound frequency value;
$\varphi$,~sound phase;
ILD,~interaural level difference;
IPD,~interaural phase difference; 
ITD,~interaural time difference;
JND,~just noticeable difference, also difference limen, or difference threshold;
$K(.)$, $K_{\mathrm{C}}$, $K_S$, $K_X$; $A$, $B$, $C$,
 $\dots$~proportionality constants; 
$l$,~sound level, also rate level of point process;
{LSO},~lateral superior olive;
{MSO},~medial superior olive;
$R(.)$, $R_{\mathrm{VS}}$,~VS,~vector strength;
$R_F$,~firing rate;
$\sigma$,~standard deviation;
\def\timijitt{t_\mathrm{J}} 
$\timijitt$~timing jitter;
$t$, $\Delta t$,~time, time difference;
$T$,~$T_{\mathrm{X}}$, ~$T_{\mathrm{(.)}}$, sound periods, time constants. 


\section{Introduction}
\renewcommand{\familydefault}{\sfdefault}
\large

Mammalian sound localization circuits contain two nuclei in the
auditory brainstem, the medial and the lateral superior olive, MSO and
LSO. Neurons in these nuclei are the first binaural neurons in the
auditory pathway connected to both ears.
Due to the physical nature of the binaural sound, the MSO neurons
process spike timing differences, Interaural Time Differences, ITDs,
in the range of tens of microseconds, 10 $\mu$s \cite{Laback2008}.
%
The MSO processes low frequency sounds, in human this is from 20~Hz to
not more than 2~kHz, and the LSO processes high frequency sounds, in
human this is from 1~kHz up to 20~kHz \cite{Middlebrooks1991}. The LSO
uses cues of the Interaural Level Difference, ILD.
The overlapping region is known to have a sensitivity drop at 1.5~kHz
\cite{Mills1958}.

\comment{ 
The highest known spike timing precision that has been proven to be
utilized in the mammalian auditory system is the motor response in a bat.
In behavioral experiment,
the big brown bat,
\textit{Eptesicus fuscus}
catches tidbit larvae of the meal-worm beetle
\textit{Tenebrio molitor}.
The motor precision is in the range of hundreds of nanoseconds,
100~{n}s, and auditory separation of ultrasound echoes in the bats
sonar sense is in the order of microsecond units, 1~$\mu$s
\cite{Simmons1998}. SHOULD BE MOVED TO DISCUSSION.
} 

Firing rate, first spike latency and individual spike timings are used
in neural system coding, especially in the auditory pathway.
Performance of the human MSO is in the range of
microseconds, 10~$\mu$s and is reported to be improved two- to
five-fold after several hours of training \cite{Middlebrooks1991}.

Different neural mechanisms are employed in the two nuclei. It has
been reported that computation in the MSO is independent on sound
intensity \cite{Grothe2010}. With higher sound intensity, first spike
latency is shortening. Relation of this dependence to ITD and ILD has
been described, yet it is difficult to interpret.
\comment{ 
The LSO and MSO extract localization information with the use of different
physical cues. The sensitivity of the system in dependence on the main
sound frequency to sounds of different frequencies should be
different. THIS IS A CRY IN THE DARK.}

In contemporary human the MSO is the larger of the two nuclei and
contains approximately 10000 - 11000 neurons and the human LSO
contains 5600 neurons \cite{Moore2000}. To implement the
loudness change
is much simpler than to record and implement
microsecond time delay. Therefore in sound generation and processing,
most of current auditory technology works as if the more important of
the two localization cues in
\textit{Homo sapiens} were the sound intensity cue \cite{Vencovsky2016}.

\comment{ 
First look at the data enables us to generalize when we listen to
harmonically and Doppler distorted sounds and when we attempt to
localize these sounds.
Speeding up, or slowing down the system time can be extended to two
octaves up and down the regular time.
Table \verb|~\ref{tab:speed-up}| has been deleted.
The range of sound frequencies processed in the MSO circuit is limited
to the low frequency band. In human, where audible sound frequencies
range from 20~Hz to 20~kHz, this low frequency band spans from 20~Hz
to 1.5~kHz.  Moreover, in the marginal values of this interval,
performance of sound azimuth discrimination is worse.
} 

It is generally agreed that the main reason, why the workings of the
MSO circuit deteriorate towards higher frequency is lowering of the
synchronization of spike trains in the circuit with the sound source
phase. The synchronization between two corresponding series of point
events can be expressed as a discrete formula of vector strength,
defined below in equation \eqref{eq:M-VS}.

This article presents description of information encoding and neural
computation in the MSO obtained mostly with analytical
computations. Using the analytical tools we extend quantitative
results obtained by numerical computations in \cite{Sanda2012}.  We
compare this analytical MSO description to the LSO description in
\cite{Bures2013} to arrive to unified description of neural circuits
in the superior olive. We use this description to find maximum spike
timing precision.
Apparently, low and high frequency sound localization use different
neuronal mechanism, because low and high sound frequencies are encoded
by distinctive codes. In a simplified view, low frequencies are
encoded by both spike timing and tonotopic organization, and high
frequencies are encoded solely by the tonotopic organization.

\section{Methods}

\subsection{Preliminaries}


In the neural circuit model used here
spikes\lb or action potentials\rb fired
in the arbitrarily precise time are individual events of
neural computation. Arbitrary precise timing would imply arbitrary
high information content. In the model, this is limited by assumptions
of intrinsic noise content. Numerical implementation has been
described in \cite{Sanda2012}. Here we develop combined stochastic and
analytical description of the model. Our aim is to arrive at
parameters and constants useful in the MSO description.

The neural circuit consists of neurons, functional units exchanging
spikes. Incoming sound is sequentially processed in the auditory
periphery. All the processing stages, including cochlea, are modeled
as all-or-none units with various degree of biological realism. After
cochlea, individual neurons correspond more-less to anatomical neuron
numbering \cite{Kulesza2007}, where the zero order neurons is the
whole mechanical-to-electrical cochlear mechanism, neurons of order 1
are in auditory nerve, order 2 are neurons in the cochlear nuclei,
order two and a half is the medial nucleus of trapezoid body\lb we
regard this nucleus as a ''polarity inverter''\rb and order 3 are the
neurons of the MSO itself. This is the binaural part of the circuit.

Before they converge on the MSO the two\lb left and right\rb branches
process sound from left and right ear, Figure~\ref{fig:circuit0}.
After the sound is encoded by cochleas
into spike trains, the rest is the processing of spike trains by
neurons. The spike trains are subject to delays and synaptic relying. 
A remarkable property of the auditory pathway is that both synapses
and neurons have the shortest response times and highest time
precision in the mammalian brain. If the neurons were represented by
R{C} circuit, or similar equivalent biophysical models, their time
constants would be comparable to, or lower than 1~ms.
Due to vernier mechanisms known from various parts of peripheral
sensory pathways, they can capture time events in the range of tens of
microseconds.
This capability has been described in human, \cite{Mills1958}.
In some animal specialists, some responses are
in the range of tens of nanoseconds, as
it has been shown in experiment on bats by \cite{Simmons1998}.
Several other time constants and frequencies are characteristic
for this neural circuit.
They are shown in Table~\ref{tab:parameters}.


\subsection{Model of the MSO neural circuit}

\begin{figure}[ht] 
\begin{centering}
\includegraphics[width=0.75\textwidth]{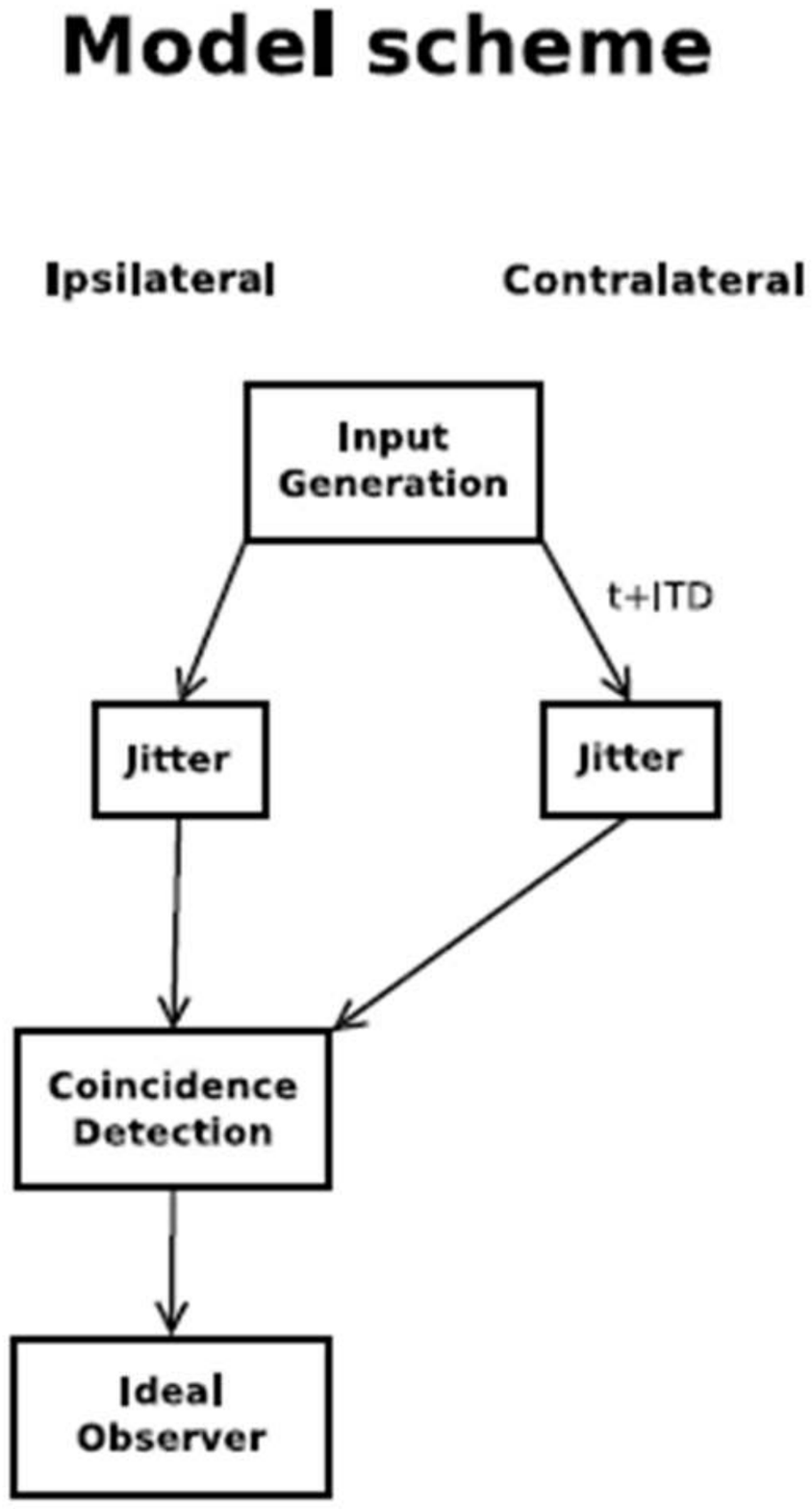}
\par\end{centering}
\noindent
\caption{ 
  \textbf{Schematic MSO circuit.}
This Figure is part of an inset reprinted with permission from the
article of \cite{Sanda2012} to show the numerical model key features.}
\label{fig:circuit0}
\end{figure}

Our model MSO circuit is based on connected phenomenological
neurons. Input sound to left and right sides is transformed by the
auditory periphery module into spike trains. Spikes in these trains are
point events, only spike times matter and the details of spike numerical
implementation do not make any differences in model output.
These spike trains
converge and diverge into higher order neurons. They are relayed
from the auditory
nerve and cochlear nucleus through the medial and lateral nuclei of
trapezoid body up to the neurons of medial superior olive, which are
first binaural neurons. Output of these neurons is the azimuth signal
encoded in a spike train. See Figure~\ref{fig:circuit0}.

\comment{ 
Schemes of the MSO circuit anatomy and the interpolation model. The
anatomical scheme (left) shows branches of the binaural circuit (only
one half of the symmetrical circuit is shown). L1, L2. $\dots$, R1, R2 are
left and right neurons. The direction from the auditory periphery to
the auditory center is oriented top-down. Each neuron is denoted by
its nucleus abbreviation (expanded in the left-hand column). Synapses
are either excitatory (E) or inhibitory (I). The model scheme (right)
employs neural circuit simplifications without losing the
implementation of the key mechanisms.
} 

\subsection{ITD readout curve}

Let us have
a monotonous function with firing rate as an input, which outputs
azimuth. We will call it the ITD readout curve. In the paper by
\cite{Sanda2012} this curve is constructed by curve fitting to
simulated points. Here we construct the curve based on assumption
that the main frequency of sound input exist, is unique and is
known. In addition to this known frequency, other parameters of the
readout curve are set to make the fitting well posed and to obtain
correct position of the curve maximum.

\subsection{Vector strength}

The vector strength
has been first used in the context of sound localization
by \cite{Goldberg1969}. Its definition follows.
Let us have sample spike phases
$\varphi_i$, $i=1, 2, \dots, N$ 
relative to phases of a given input master periodic function, which
does not enter the formula. The periodicity of tones making up speech
is a perfect example of such stimulus.
Discrete sum vector strength of sample
\( \varphi_1, \dots, \varphi_N \) attains values from 0 to 1 and
is defined as
\begin{equation}
R_{\mathrm{VS}}(\varphi_i) = \frac{1}{N}
\sqrt{\left( \sum_{i=1}^N \cos\varphi_i \right)^2
+ \left( \sum_{i=1}^N\sin\varphi_i \right)^2}.
\label{eq:M-VS} 
\end{equation}

\subsection{JND and ideal observer}

A higher variability of firing leads to a lower precision of the rate
code. Intuitively, if a repeated presentation of the same stimulus
evokes each time different spike count, then to distinguish between
two different stimuli, the associated spike count change must be larger
than the spike count variability.
This way we determine the Just Noticeable Difference (JND) of the rate code.
In other words, this is the precision of rate coding.

We will ask whether it is possible to distinguish between two random
processes with rates $l_{1}$ and $l_{2}$, $l_{2}>l_{1}$. If we count
events in a given counting window, we get counts $n_{1}$ and
$n_{2}$. The probability that the observer obtains a result that
$l_{2}>l_{1}$ equals to the probability that $n_{2}>n_{1}$. Let us
assume that the random variables $n_{i}$, $i=1,2,\dots$, have
probability distributions $p(n_{i})$ with means $\mu_{i}$ and equal
standard deviation $\sigma$. A detection distance is then defined
\cite{Tanner1961} as
\begin{equation}
d'=\frac{\mu_{2}-\mu_{1}}{\sigma}.
\label{R_SDT1} 
\end{equation}

This definition expresses the fact that the larger is the variance of
the spike count, the worse is the detection capability.
In psychophysics, a threshold value is commonly defined as that value
for which the percentage of correct answers equals 75\%. In our case,
the examined value is the just-noticeable change of firing rate,
$\Delta l=l_{2}-l_{1}$.
Assuming that both $p(n_{1})$ and $p(n_{2})$ are Gaussian (normal) distributed,
the 75\% probability of $n_{2}-n_{1}>0$ corresponds to $d'= 1$. To
obtain the JND of firing rate, we scale the detection distance with
$\Delta l$ and put $\delta'=d'/\Delta l$. Then, the JND of firing rate
is
\begin{equation}
\Delta l_{\mathrm{JND}}=1/\delta'=
\frac{l_{2}-l_{1}}{\mu_{2}-\mu_{1}}\sigma.
\label{R_SDT2} 
\end{equation}

\section{Results}

We investigate how the MSO circuit output and overall performance depend on
sound frequency and sound intensity.
Figures from \ref{fig:FM1V1} to \ref{fig:FM4V1} show numerical simulations
and experimental data as black lines and points, analytically described
lower estimates as blue lines, upper estimates as red lines and
standard errors of measurement and variations are shown in green.


Successive figures show individual steps of sound localization
processing. Fig.~\ref{fig:FM1V1} shows how vector strength in
individual units lowers towards high sound frequencies.
Fig.~\ref{fig:FM2V1} shows the range of ITDs in sound localization judging.
Fig.~\ref{fig:FM3V1} shows the JND of the neural circuit in the
dependence of the spike timing jitter.
Fig.~\ref{fig:FM4V1} shows synchronization to main sound
frequency (in the case when it exists).

Figure \ref{fig:FM1V1} shows how vector strength $R_{\mathrm{VS}}$
lowers towards higher frequencies,
as it can be observed in the module of the auditory
periphery consisting of auditory nerve and cochlear nucleus. 
The prevailing
majority of neurons in the auditory pathway has vector strength
spike train statistics sigmoidally dropping towards higher sound
frequencies as it is in this example. In this figure, data
originally recorded by \cite{Joris1996} at the MSO of domestic cat,
were fitted to the sigmoidal curve with the general formula of the
Boltzmann function used in \cite{Marsalek2005}.
The curve fit of vector strength $R_{\mathrm{VS}}$ in dependence on sound
frequency $F_\mathrm{S}$ is:

\begin{equation}
R_{\mathrm{VS}} =
1/ (1+\exp(K_{\mathrm{S}} f_{\mathrm{S}} -K_{\mathrm{C}} F_{\mathrm{C1/2}} )),
\end{equation} 
where parameters with values are
$K_{\mathrm{S}}= 2/0.75 = 2.666$~ms~ (kHz$^{-1}$), sound frequency coefficient; 
$K_{\mathrm{C}}= 4/0.75 = 5.333$~ms~(kHz$^{-1}$), critical coefficient;
and $F_{\mathrm{C1/2}}= 0.75$~kHz, critical half frequency.
The numerical values are the proportionality constants of the
$R_{\mathrm{VS}}$ upper bound.
Note that at sound frequencies from 20 to
100 Hz, there are two branches reflecting the existence of two
alternative ways how to compute the lower limit, which is 90 \%
of the upper limit, see Discussion section.

Figure \ref{fig:FM2V1} shows the curves limiting the ITD
obtained with the basic parameter set in dependence on the
sound frequency.
The quadratic curve fit of the JND denoted $\Delta t_{\mathrm{JND}}$ is:
\begin{equation}
\Delta t_{\mathrm{JND}}
= A (f_{\mathrm{S}} - F_{\mathrm{C1/2}})^2 + B,
\end{equation}
%
Values of these parameters are $A=10^{-5}$, $B=0.05$, 
$F_{\mathrm{C1/2}}=$~1~kHz.
The parameters in the Figures were constructed as follows:
firstly, splines were fitted to experimental
data. Splines parameters were allowed to vary within the 10 \%
of their original value and were approximated by the rounded off
values. These procedures were also used to get parameter values below.
For ranges of the audiogram parameters, see
also \cite{Zwislocki1956}. Analogously to Figure \ref{fig:FM1V1},
sound frequencies from 20 to 100 Hz exhibit higher spread between
lower and upper limits, as the fitting method used\lb quadratic fit\rb
is the same for both limits.

Figure \ref{fig:FM3V1} shows how the
JND of ITD depends on timing jitter magnitude $\timijitt$.
Figure purpose is to capture, what is the best JND.
There are several time constants, which are defined in relation to
physical properties of spatial sound processing.
To attain to rounded off parameters as in the other
figures, we select individual values of the spike timing jitter and
describe their purpose in the localization precision estimation.
{\it Critical} timing jitter is lower estimate of timing jitter
captured by spike train of typical mammalian neuron,
$T_{\mathrm{JC}}~=~0.2$~ms.
{\it Normalized} timing jitter $T_{\mathrm{JN}}~=~1$~ms
is the value of timing jitter normalized in relation to the output JND
with respect to average firing rate.
{\it Optimal} timing jitter $T_{\mathrm{JO}}~=~1.66$~ms 
is a result of intersecting two fits described below.

Simulations show that with lowering
timing jitter the circuit output is virtually more and more precise.
Yet, when the jitter is lower than critical value $T_{\mathrm{JC}}$,
determined by intrinsic noise, duration of coincidence detection window,
and by other time constants, the precision lowers again.
The two curves fitted to the simulation are:
1. fit of exponential function to simulations, red curve:
\begin{equation}
\Delta t_{\mathrm{JND}}= \exp(A_1(\timijitt-B_1)) - C_1,
\label{eq:exponential}
\end{equation}
where $A_1=1.9$, $B_1=1.25$ and $C_1=0.2$ are fitted parameters. This
relation is shown conveniently by the logarithmic y-axis in this Figure.

2. another fit, which also takes into account shot noise in lower
jitter values, is to a quadratic function, blue curve,

\begin{equation}
\Delta t_{\mathrm{JND}} = A_2(\timijitt-B_2)^2 + C_2,
\label{eq:parabolic}
\end{equation}
where $A_2=2.5$, $B_2=1$, $C_2=1$. There is only one parameter sought
by numerical simulation. This is $A$, fitted to data, as the point
$(x,y) = (B_2,C_2)$ has been chosen to be a unit. This fit is the
normalized fit of the model.

Logarithm of the simulated JND lowers with the exponential curve
\eqref{eq:exponential}, which is concave function of sought jitter
$\timijitt$
as the jitter gets lower. The trend towards higher
accuracy diverges from the parabolic fit in equation \eqref{eq:parabolic},
when jitter reaches critical value between
$T_{\mathrm{JC}}$ and $T_{\mathrm{JO}}$.

\begin{equation}
T_{\mathrm{JC}}=0.2
\lessapprox
\timijitt
\lessapprox
T_{\mathrm{JO}}
\lessapprox
2\mathrm{ms}. 
\label{eq:jitters}
\end{equation}
Beyond that point towards the lower jitter values, the neural circuit
cannot function properly, as too low jitter prevents the interaction
of spikes from the left and right side within the coincidence
detection mechanism.

This corresponds to the analytical dependency obtained in \cite{Salinas2000}
for a perfect integrator model with several
inputs. The mechanism studied thereof is close to the MSO neural
mechanism studied here.
The firing rate changes in dependency on the input spike timing
variability of {\it partially correlated}
input spike trains.

Figure \ref{fig:FM4V1} shows the ITD readout curve. 
The rising slope
of this curve is used as a read{out} function yielding the firing rate
in dependency on the ITD, which in turn signals the sound azimuth to
the next nuclei of the auditory pathway.

In the numerical model of \cite{Sanda2012} we reproduced a procedure
to obtain azimuth tuning curves. 
In this procedure, 
prior assumption was the existence of the main sound frequency.
%
When we use this assumption the result is less general than in the numerical
model, yet we obtain a fitting curve which is more coherent. (It has the
higher vector strength value.)
Two estimate errors are present in this Figure. The first is the
mismatch between the use of Gaussian (normal) probability density
function, as it is used in the experimental literature, and a circular
statistics.
More details of the circular statistics use in the sound localization
context are explained in the article on the ergodic assumption by
\cite{Toth2018}.
The first error is shown here as the green curve.
The other error, not shown
in Figure, would arise from the 
prior assumption above. It appears that the assumption of the main
frequency existence leads to more precise estimates.

\begin{table}
\begin{centering}
\begin{tabular}{|l |l | r| c | c |}
\hline 
Parameter & Symbol & Units & Typical Value & Ranges\tabularnewline
\hline 
\hline 
Timing Jitter & $\timijitt, \sigma$ & ms & 1 & 0.125 - 8 \tabularnewline
\hline 
Window of Coincidence Detection &
$w_{CD}$ & ms & 0.6 & 0.15 - 1.5 \tabularnewline
\hline 
Sound Frequency & $f$ & Hz & 200 & 40-1600 \tabularnewline
\hline
Shortest Perceptual Time &
$T_{PT}$ & ms & 20 & 20 - 80 \tabularnewline
\hline
\end{tabular}
\par\end{centering}

\caption{\label{tab:parameters}
The basic set of parameters. 
}
\end{table}


\section{Discussion}

From the time of the mechanistic description of the house fly
``hard-wired'' motion detector by \cite{Reichardt1961} and
``hypothetical'' sound localization circuit by \cite{Jeffress1962},
there have been numerous models capturing essential workings of neural
circuits. Based on psychophysical measurements, models of the MSO
neural circuit, like \cite{Bures2012}, \cite{Franken2014},
\cite{Grothe2010}, \cite{Joris2006}, \cite{Marsalek2005},
\cite{Toth2015}, arrived to the ``canonical'' set of parameters
describing this neural circuit and connected nuclei in the auditory
pathway.

In this paper we have revisited numerical simulations by
\cite{Sanda2012}. We have added analytical estimations to the description
of the MSO function, which have not been known previously.
Our analytical calculations make possible to derive
time constants useful in description of normal human
hearing. The descriptions are valid also for
hearing with hearing aids and cochlear implants.
All Figures contain analytically expressed upper and lower limits in
their transfer functions or other functional descriptions.

\begin{figure}[ht] 
\begin{centering}
\includegraphics[width=0.75\textwidth]{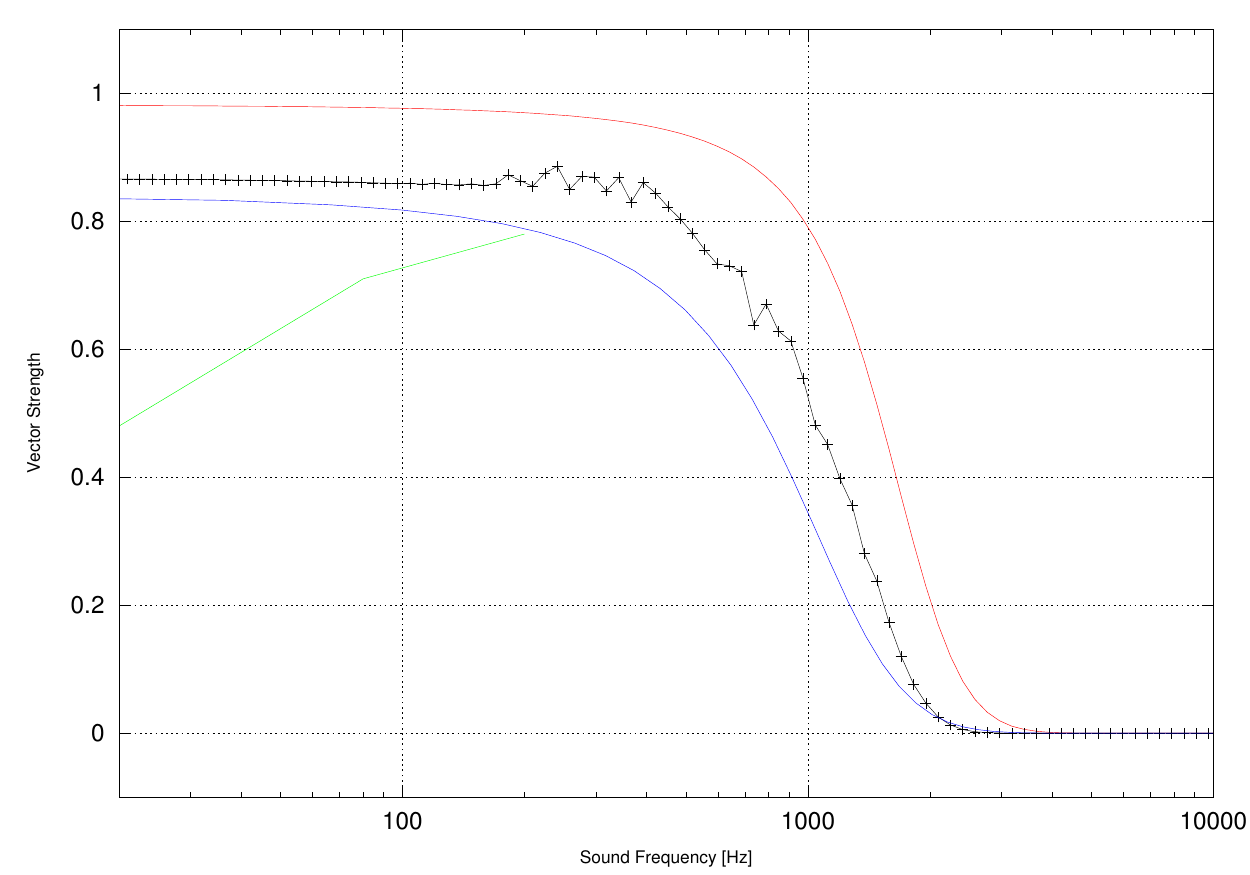}
\par\end{centering}
\noindent
\caption{ 
\textbf{Vector strength of auditory nerve spike trains
in dependence on sound frequency.}
X-axis shows sound frequency in~Hz in logarithmic scale
and y-axis shows the vector strength.
Even though in some nuclei up the auditory pathway the synchronization
can be maintained towards higher frequencies than shown here, the
decrease of the vector strength towards higher frequencies is a
general property of all neurons in the auditory pathway.
Red curve shows upper theoretical limit, blue curve shows lower
limit, green curve shows limit imposed by lower firing rate
and lower energy of low frequency sound.
Black points are data simulated with the use of
point-process spike train generation with the use of the dead time
Poisson process.
Note that in frequency $f_\mathrm{S}$ range from 20 to 200 Hz the
lower limit is shown by curve branching to two branches to the
left. The upper is the Boltzmann function fit and the lower is
decrease of vector strength at low frequencies due to stochastic
response of high spontaneous rate neurons.}
\label{fig:FM1V1} 
\end{figure}

Figure \ref{fig:circuit0} shows workings of the model neural circuit
as simple scheme.
Figure \ref{fig:FM1V1} contains two lower limit branches at low
frequencies (shown by the blue and green curves, respectively).
The green curve uses an assumption of lower energy and lower
contribution to spike rate in neural units in lower frequencies.
Limiting lower bound by two different analytical functions (branches)
can be understood as the estimate uncertainty.
A conservative estimate of the lower bound always considers the lower
of the two branches.
This uncertainty should be recognized as one of original results
presented for the first time in this paper. Its existence has been
proposed in a doctoral thesis by \cite{Bures2014}. To our knowledge
this observation has not yet been published elsewhere.

\begin{figure}[ht] 
\begin{centering}
\includegraphics[width=0.75\textwidth]{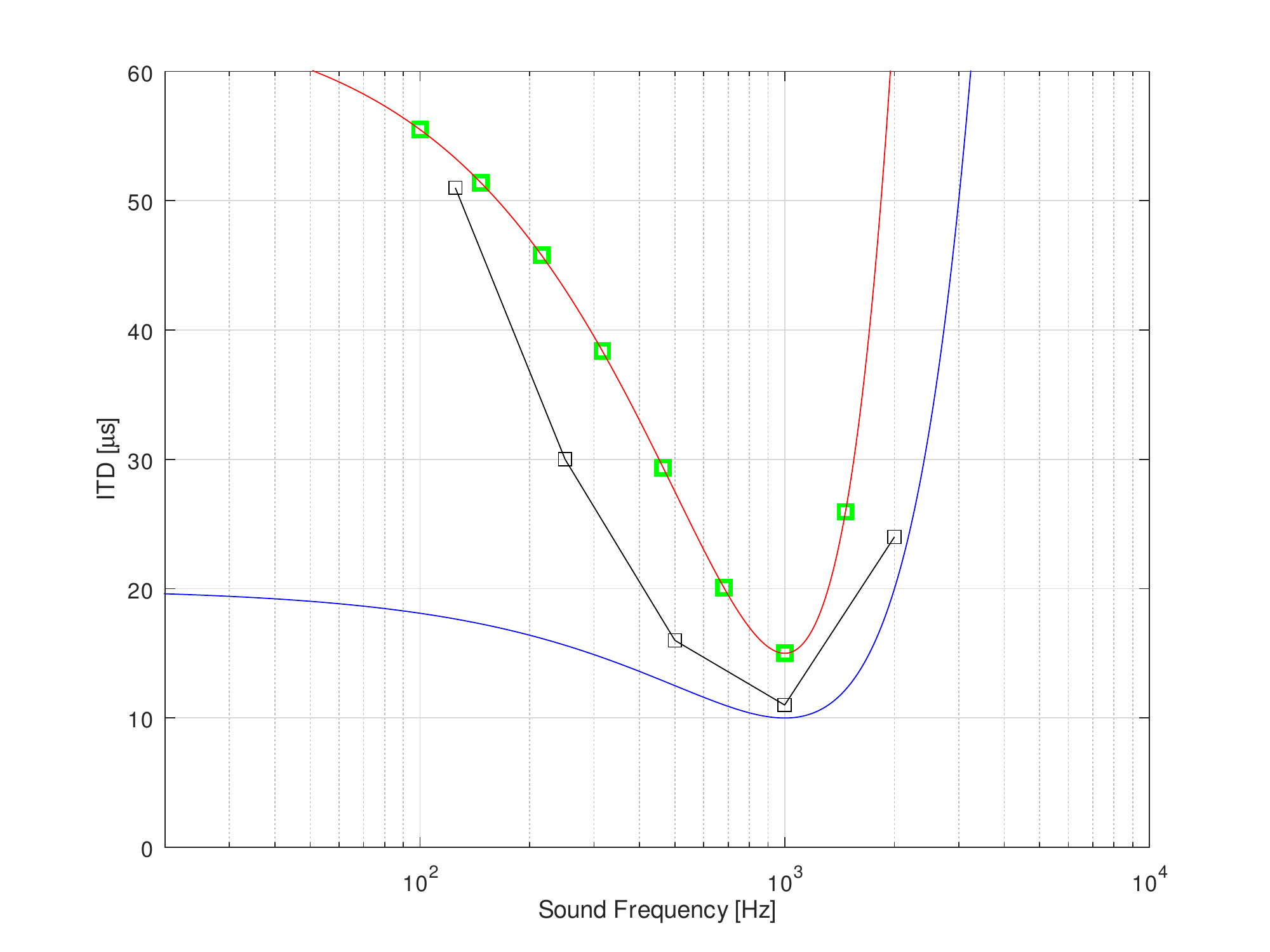}
\par\end{centering}
\noindent
\caption{
\textbf{The shortest JND of ITD detected in the dependence on sound frequency.}
X-axis shows sound frequency in~Hz in a logarithmic scale and y-axis
the shortest JND of ITD in $\mu$s. This is a theoretical prediction
based on the analytical model and basic parameter set used in
simulations. As in other figures, black line is obtained by simulation
and red and blue lines are respectively upper and lower bounds
obtained by an analytic fit.
}
\label{fig:FM2V1} 
\end{figure}

Figure \ref{fig:FM2V1} depicts a quadratic fit.
Clearly the data cannot be captured by the linear curve.
The procedure to obtain the fit is analogous to obtaining other
parameters in this paper. Initially splines were used and then their
output was rounded to arrive to the quadratic fit.
This fit is the simplest analytical way, how to capture nonlinear
and band limited span of human hearing range.

\begin{figure}[ht] 
\begin{centering}
\includegraphics[width=0.75\textwidth]{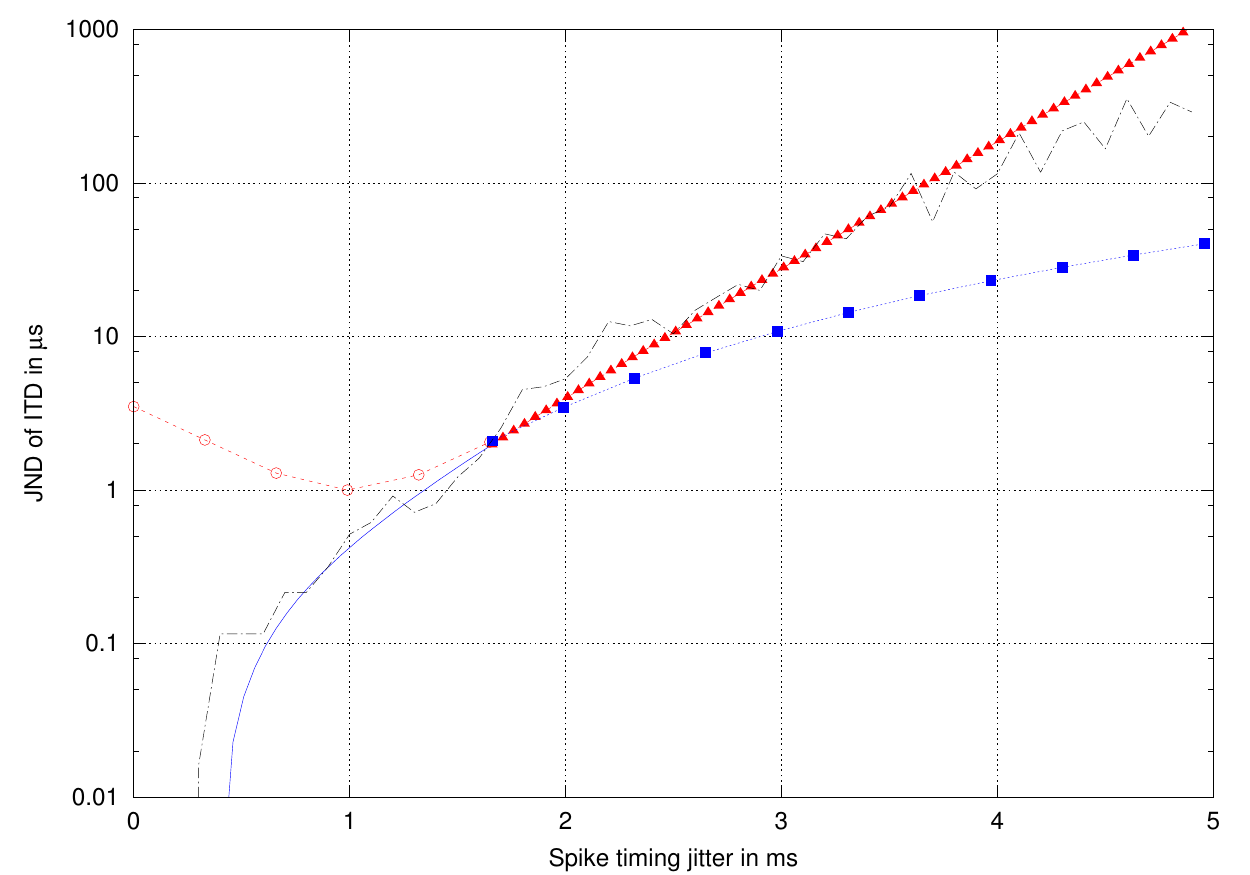}
\par\end{centering}
\noindent
\caption{
\textbf{JND values in basic parameter set
in dependence on the spike timing jitter magnitude.}
This plot in semi-logarithmic y-scale shows JND (just noticeable difference)
of interaural time difference depending on variation of the spike
timing jitter. Jagged black line: simulated data,
solid line: an exponential fit to the simulations under the assumption
of arbitrary time precision in the model circuit, 
dotted line: a quadratic function estimate of spike timing precision
in a system with addition of noise.
Note that in this Figure 
the exponential and quadratic fits cross at $f_\mathrm{S}$ = 1.66 ms.
In order to correspond to other Figures showing the upper and lower
bounds of the estimate of stochastic model, the two fits are split
into two branches of the same function at this point of
$f_\mathrm{S}$ = 1.66 ms.
For lower x-axis $\timijitt$ values, quadratic fit is larger than the
exponential, and vice versa. This is indicated by distinctive
data-points.
(These are circles and triangles; no data-points and squares, respectively.)
Also notice that the curve of the quadratic fit goes through the point
[1,1], this is a consequence of using normalized parameter set.
}
\label{fig:FM3V1} 
\end{figure}

Figure \ref{fig:FM3V1} calculations use assumptions about intrinsic noise
\cite{Bures2014}.
The simulation data have been obtained by arbitrary precision
calculation. Any neural recording cannot reach this precision due to
the internal noise of both neurons and recording electronics. In order
to capture circuit noisiness, we have used both exponential and quadratic
fits. When we attempt to use them as upper and lower bounds, we notice
that they exchange their order in region close to the optimal jitter
value.
In other words, at the lower jitter values the two estimates exchange
their ordering.
This is the choice of the quadratic fit to obtain a normalized bound
together with other data-points. Numerical simulation with the basic
set of parameters around the x-axis value of $\timijitt$ = 1 ms lies
beyond this point, but close to the exponential fit.
%

\begin{figure}[ht] 
\begin{centering}
\includegraphics[width=0.75\textwidth]{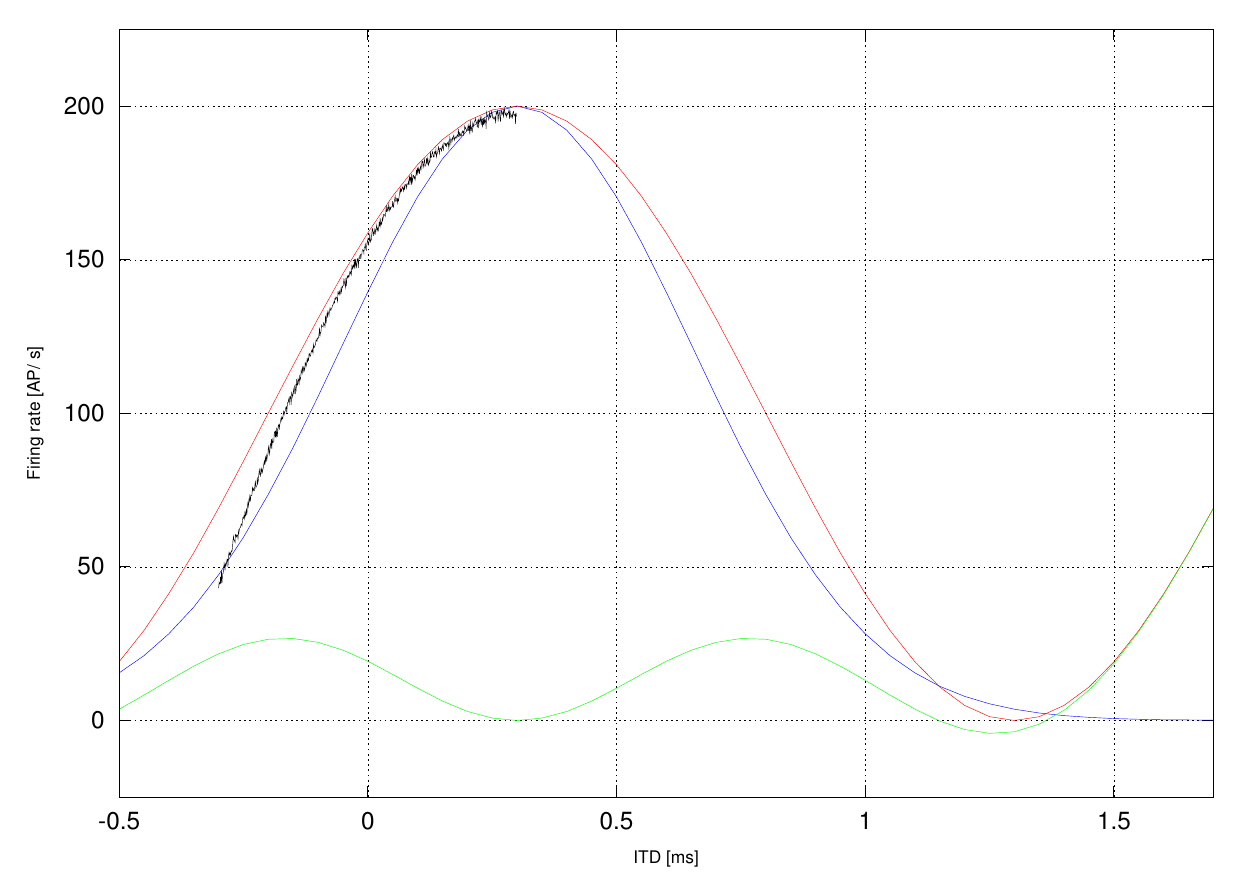}
\par\end{centering}
\noindent
\caption{
\textbf{Fit of example functions to firing rate slope ITD readout curve.}
X-axis shows ITD in~ms and y-axis shows corresponding firing rate in
action potentials per second. Note that the curve peak is offset from
the origin of coordinates at $t_{\mathrm{ITD}}=0$. Red curve is the Sine
density function fit and blue curve is fit by the normal density
function with the variance set to correspond to the known sound main
period. Green curve shows the difference between the red and blue
curves.
}
\label{fig:FM4V1} 
\end{figure}

Figure \ref{fig:FM4V1} contains better fit of the Sine function, as
compared to \cite{Sanda2012}.
As the lower bound we can also use circular normal density function,
the difference is negligible, not shown. The conditions, when it is
possible to interchange the estimated probability density functions of
regular normal (Gaussian) and circular normal densities are discussed
in detail in \cite{Toth2018}.
Comparison of time constants and sound periods in the model presented
here will answer a tentative question: What is the highest slope of
the ITD interpolation curve, such that it gives the resolution of
well known minimum audible angle in the midline (ITD = 0), which is
4$^\circ$ in angular degrees? This slope is
more steep in higher frequency sounds, its maximum is attained in
maximum frequency of the MSO circuit operation, around 1~kHz
\cite{Marsalek2005}.

Towards the analytical descriptions it is important to note that other
periodic functions can be used as the ITD readout curves. In
\cite{Toth2018} we have compared the
\textit{Sine function} with the \textit{circular beta density},
\textit{circular normal density},
and other alternative functions. To impose periodic and
infinite boundaries to the problem, regular normal density and
circular normal density have been used and tested in previous versions
of our model. No differences between these densities with proper
parameters have been shown by common statistical tests when testing
differences between two probability densities, ibidem.

\cite{Franken2014} use recorded spike trains of several nuclei in the
MSO neural circuit to demonstrate that coincidence detection is an
essential part of the neuronal arithmetic executed by the neural
circuit.
These authors show simulations combined
with experimental description of MSO workings in line with
findings of this paper and with coincidence detection theories.
Another MSO model, already studied in 2005 is: \cite{Zhou2005}, this
is an example of simplistic model, motivating the neural circuit description
presented in this paper.
For discussion of neural coding in the auditory nerve, auditory pathway,
cochlear implants and brainstem neural circuits see \cite{Kerber2012}.

Our investigation of quantitative properties of the superior olive
neural circuit is also motivated by the three LSO experimental papers,
which have detailed methodology applicable to LSO, to the overlap of
sound frequency ranges between the LSO and the MSO; and also to the
MSO range itself;
\cite{Joris1995}; \cite{Joris1996}; \cite{Joris1998_envelope}.

%
Following his
paper from 1948, Lloyd A. Jeffress
dedicated lots of efforts to the search of a mechanism, by which
microsecond time scale events of directional sound difference can be
transformed into a code processed and transmitted by action potentials
lasting several microseconds \cite{Jeffress1962}. 
Historical comments on Jeffress papers are summarized by
Cariani in Scholarpedia \cite{Cariani2011}. A plausible
explanation of the microsecond precision of the MSO circuit describes
the neural computation by leading edges of action potentials and
post-synaptic potentials, \cite{Marsalek2000}, \cite{Toth2018}.

%

In \cite{Marsalek2000},
individual steps of signal processing in the superior olive neural
circuits have been investigated. Various synaptic mechanisms have been proposed
\cite{Marsalek2005_Kofranek}.
Spike timing jitter and spike variability have been
systematically analytically investigated by \cite{Kostal2010}.
%

Article by \cite{Michelet2012} discusses interaural phase delays
(IPDs; when they exist, their utility is equivalent to that of the
ITDs) and cochlear delays.  For cochlear delays, very important is to
review the ranges of delays in comparison to sound periods and
classically described excitatory-excitatory and excitatory-inhibitory
responses to binaural inputs in \cite{Joris2006}.
Paper of Srinivasan, Laback and Majdak cites current progress of ITD
encoding by binaural cochlear implants, this is important for model
validations and applications to studies with hearing aids and
electrical hearing, \cite{Srinivasan2018}.



\subsection*{Conclusions}

This theoretical paper is continuation of sound localization precision
descriptions in the MSO \cite{Sanda2012} and in the LSO
\cite{Bures2013}.
Major novel results here are following:
1) analytical estimates of results obtained previously only by
numerical simulation and
2) estimates of auditory parameters using functions bounding them from
the lower and from the upper values.
These are known characteristics of the human sound
localization circuit.

\subsection*{Summary}

This pre-print is written as a sequel to \cite{Sanda2012} and
\cite{Bures2013}. \cite{Sanda2012} presents numerical simulation of the
MSO circuit. \cite{Bures2013} argues that the putative ``spike rates
subtracting'' neural computation in the LSO must consist of
coincidence detections confined in short time and narrow frequency
band. In this paper we recalculate all the parameters of the MSO model
analytically, where possible. We also show, how the parameters are
bound from lower and upper sides.

\subsection*{Acknowledgments}

This project was in part funded by
Charles University graduate
students research program, acronym S{V}V,
No. 260 519/ 2020-2022, to Petr Marsalek.
{Complete sources are available upon request at
the following e-address: Petr.Marsalek At LF1.CUNI.CZ}

\comment{ 
We acknowledge uploading this manuscript first to
Ar-Xiv repository prior to impacted journal submission.
Special thanks to
Marek Hajny
for reading and commenting the manuscript.
} 

\subsection*{Author contributions}

Model design: PM, Z{B}.
Methodology: PM, PS.
Writing original draft: PM.
Writing and editing: PM, PS, Z{B}.

\comment{ 
\medskip
\subsubsection*{Technical, \TeX\ Typesetting, Archiving Notes}
\small
\noindent
{Compilation notes:}
\LaTeX\ compilation date is \today.
\noindent
{Bib\TeX\ output:} Circa 30 is the estimated
references count (March 2020: 32).
Version of sources in octave is 2020.1.4,
version of manuscript is 2020.A.2.
{Complete sources are available upon request.}
{The sources web link /protected by a password/ is here.}
{\tt http://nemo.lf1.cuni.cz/mlab/ftp/ftp/bures-z}\par
\noindent
{Preview web link:}
{\tt http://nemo.lf1.cuni.cz/mlab/ftp/ftp/to-be-added}
\par \noindent
Things to do to by co-authors:\par
\noindent
Petr Marsalek. Clean up and finish first pass writing.\par
\noindent
Pavel Sanda is kindly asked to help the author team with tasks:
1) reading Discussion 
2) checking up figures. 
Zbynek Bures: should do internal English language copy-editing.
final reading of the manuscript before
submission to the journal. 
} 


\bibliographystyle{apalike} 

\bibliography{../sound_local}


\end{document}